# Janus Waves

**DIMITRIS G. PAPAZOGLOU,[1,2*] VLADIMIR YU. FEDOROV,[3,4] STELIOS TZORTZAKIS [1,2,3]**

[1] *Institute of Electronic Structure and Laser, Foundation for Research and Technology-Hellas, P.O. Box 1527, 71110, Heraklion, Greece*
[2] *Department of Material Science and Technology, University of Crete, P.O. Box 2208, 71003, Heraklion, Greece*
[3] *Science Program, Texas A&M University at Qatar, P.O. Box 23874, Doha, Qatar*
[4] *P.N. Lebedev Physical Institute of the Russian Academy of Sciences, 53 Leninskiy Prospekt, 119991, Moscow, Russia*
*\*dpapa@materials.uoc.gr*

We show the existence of a family of waves that share a common interesting property affecting the way they propagate and focus. These waves are a superposition of twin waves, which are conjugate to each other under inversion of the propagation direction. In analogy to holography, these twin "real" and "virtual" waves are related respectively to the converging and the diverging part of the beam and can be clearly visualized in real space at two distinct foci under the action of a focusing lens. Analytic formulas for the intensity distribution after focusing are derived, while numerical and experimental demonstrations are given for some of the most interesting members of this family, the accelerating Airy and ring-Airy beams

The formation of twin images, a "real" and a "virtual" one, as light is diffracted by a hologram is a well know effect [1]. The hologram acts as a spatial modulator of the amplitude and phase of the original wavefront, and thus a complex wavefront is formed. This wavefront can be decomposed into two waves [2], one of which leads to the formation of the "real" image of the original object at some distance $z_o$ and the other to the formation of the "virtual" image at distance $-z_o$. These two inseparable waves interfere, and may deteriorate the reconstructed object. In holography the twin image effect is undesirable because of the deterioration of the images and has been successfully obviated [3] by introducing a separate tilted reference wave during the hologram recording process.

Here we show that the formation of symmetric twin images can be observed not only in holograms but also for a broad class of waves that we will call in the following as Janus Waves (JWs). Like the god Janus from the Roman mythology, who was depicted with two faces looking in opposite directions JWs can be decomposed to the propagation of two waves, which are conjugate to each other under inversion of the propagation direction. Such waves include beams like the accelerating Airy [4-6] and ring-Airy beams [7-11] and higher order accelerating beams [12-16]. A formal definition of JWs will be provided in the following. The analogy to holography holds if we consider the intensity peaks, or focus points, of these waves as an object. In this case the formation of twin images manifests as two foci a "real" and a "virtual" one symmetrically positioned to the plane of symmetry.

Interestingly this effect was, to our knowledge, not observed up to now. In contrast to classical holograms, and due to the geometry of the beam, the part of the wave that is related to the "virtual" image is diffracting out with minor interference with the part of the wave that is related to the "real" image. The presence of the twin "virtual" image clearly manifests when these beams are focused by a lens. In this case both the converging and the diverging part of the beam are brought to focus and form a pair of opposite facing focal distributions.

**Theoretical analysis**

Let's assume that a harmonic wave described by its field $U(\mathbf{r}, z)$, where **r** is the transverse position vector and $z$ is the position along the propagation axis, can be described as the superposition of two waves, $\Psi_o(\mathbf{r}, z)$ and a "conjugate" $\Psi_o^*(\mathbf{r}, -z)$:

$$U(\mathbf{r}, z) = \Psi_o(\mathbf{r}, z) + \Psi_o^*(\mathbf{r}, -z) \tag{1}$$

These conjugate waves (CW) are in perfect analogy with in-line holograms, where the diffracted wave is formed by the superposition of the "real" and "virtual" image. One can easily show that this wave is conjugate symmetric under inversion of the propagation direction, $U(\mathbf{r},-z) = U^*(\mathbf{r},z)$. Furthermore for such a field the distribution at z=0, which defines the plane of symmetry, is always a real valued function:

$$U(\mathbf{r},0) = \Psi_o(\mathbf{r},0) + \Psi_o^*(\mathbf{r},0) = 2\text{Re}[\Psi_o(\mathbf{r},0)] \tag{2}$$

The mathematical criterion for CWs, is simple and clear and is based on eq. (2): *"A wave is Conjugate if its field $U(\mathbf{r},z)$, is real valued at a transverse plane along its propagation"*. By real valued we mean that the phase can take only $m \cdot \pi$ values (where $m = 0, \pm 1, \pm 2 \ldots$). In what follows we prove the validity of the CW criterion. Without loss of generality we can set the origin of our coordinate system on a plane so that: $U(\mathbf{r},0) = \text{Re}\{U(\mathbf{r},0)\}$. The propagation of such a wavefront can be described using the concept of the angular spectrum [2]. The angular spectrum $P_o(\mathbf{k}_\perp) \equiv \iint U(x,y,0)e^{-i(k_x x + k_y y)}dxdy$, where $\mathbf{k}_\perp$ is the transverse component of the wave vector, of such a real valued field is Hermitian $P_o(-\mathbf{k}_\perp) = P_o^*(\mathbf{k}_\perp)$ and thus can be described as:

$$P_o(\mathbf{k}_\perp) = |P_o(\mathbf{k}_\perp)| e^{i\varphi_o(\mathbf{k}_\perp)} \tag{3}$$

where the amplitude $|P_o(\mathbf{k}_\perp)| = |P_o(-\mathbf{k}_\perp)|$ of the angular spectrum is an even function, and $\varphi_o(-\mathbf{k}_\perp) = -\varphi_o(\mathbf{k}_\perp)$ is an odd phase distribution in the k-space. Using the angular spectrum at $z = 0$ we can estimate the amplitude of the wave at any distance z along the propagation axis:

$$U(\mathbf{r},z) = \iint |P_o(\mathbf{k}_\perp)| e^{i\varphi_o(\mathbf{k}_\perp)} e^{iz\sqrt{k_o^2 - k_\perp^2}} e^{i(k_x x + k_y y)} dk_x dk_y$$
$$= \Psi_o(\mathbf{r},z) + \Psi_o^*(\mathbf{r},-z) \tag{4}$$

where $\Psi_o(\mathbf{r},z) \equiv \frac{1}{2} \iint |P_o(\mathbf{k}_\perp)| e^{i\varphi_o(\mathbf{k}_\perp)} e^{iz\sqrt{k_o^2 - k_\perp^2}} e^{i(k_x x + k_y y)} dk_x dk_y$.

Thus the propagation of a wave that is real valued, at a transverse plane along the propagation, can always be described as the superposition of two waves, $\Psi_o(\mathbf{r},z)$ and a "conjugate" $\Psi_o^*(\mathbf{r},-z)$.

A large variety of waves fulfill the CW criterion, ranging from accelerating beams such as Airy [4, 17] and ring Airy beams [7, 8], but also the widely used Gaussian beams and Bessel beams [18]. On the other hand Helical beams [19], are waves that do not fulfill this criterion due to their helical phase gradient.

From this broad category of CWs we now concentrate on those waves that exhibit a discrete focus away from their symmetry plane. We define such waves as JWs. These include for example all accelerating beams [4-9, 12, 13] but not Gaussian and Bessel beams [18]. What makes JWs interesting is their peculiar behavior when they are focused by a lens. As we will show in the next session, two foci, instead of one, can be observed.

A characteristic example of a JW is the ring-Airy beam [7]. These beams fulfill the CW criterion and exhibit a discrete focus. More specifically, they are real valued at their generation plane since they are described [7, 8] by $Ai(\rho)e^{a\rho}$, where $Ai(.)$ is the Airy function, $\rho = (r - r_o)/w$ and $r$ is the radius, $r_o$, and $w$ are the ring radius and width parameters respectively and $\alpha$ is the apodization coefficient. Furthermore, they also exhibit a discrete focus, away from their symmetry plane, since they abruptly autofocus [7, 8] at predefined distance $z_o$ [9].

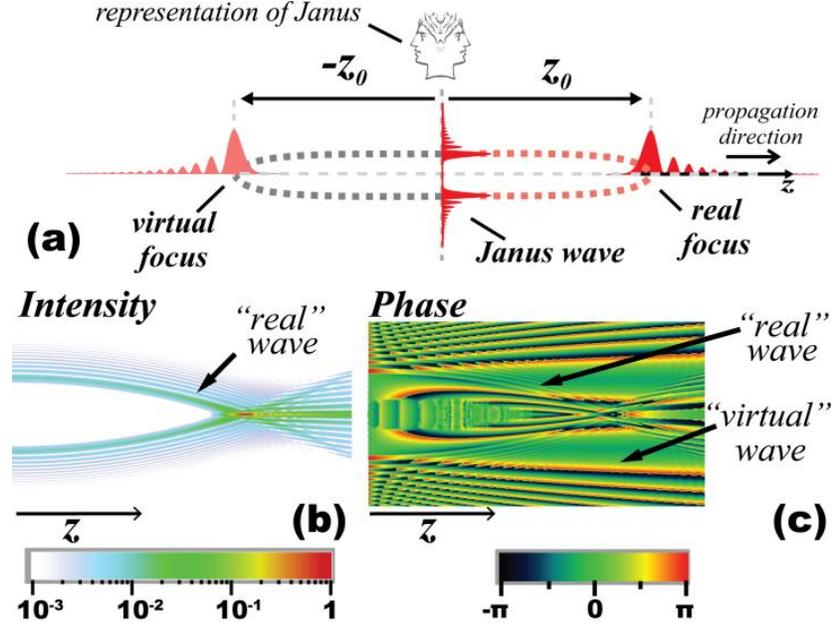

**Fig. 1**. Propagation of Janus waves as a superposition of "real" and "virtual" waves **(a)** graphical example of a JW under the form of a ring-Airy beam. (the two-faced Roman mythology god Janus is also represented) **(b), (c)** Cross sectional (*r-z*) simulation results of ring-Airy propagation. The parabolic trajectory towards the focus is clearly visible both in the intensity (logarithmic scale) and the phase (wrapped phase). The virtual wave existence although practically invisible in the intensity map is clearly visible in the phase

A graphical example of a JW propagation under the form of a ring-Airy beam is shown in Fig. 1(a). The focal point lies at a distance $z_o$ while a "virtual" focal point lies on the opposite side (at $-z_o$) and is not accessible. The effect of the twin "real" and "virtual" waves on the beam propagation can be visualized by numerically simulating the propagation of a ring-Airy beam. As shown in Figs. 1(b),(c), the characteristic parabolic trajectory towards the real focus is clearly visible both in the intensity (Fig. 1(b)), and the phase of the beam (Fig. 1(c)) as depicted by, the phase wrap uncovered, $\pm\pi$ iso-phase curves. On the other hand, a signature of the existence of the virtual wave can be seen only in the phase map. The intensity, although in logarithmic scale, shows no trace of the component that diffracts out. The "virtual" wave manifests as a diffracting out wave and is usually ignored [16]. If we could reverse propagate this diffracting part then we would observe a symmetric to the initial focus, abrupt focus.

*Janus wave focusing by a lens*

If we insert a thin lens on the plane of symmetry ($z=0$) of a JW then the wave distribution after the lens is described by [20]:

$$U'(\mathbf{r},z) = \frac{1}{\zeta} e^{\frac{-ik(x^2+y^2)}{2f\zeta}} \left\{ \Psi_o\left[\frac{\mathbf{r}}{\zeta}, \frac{z}{\zeta}\right] + \Psi_o^*\left[\frac{\mathbf{r}}{\zeta}, \frac{-z}{\zeta}\right] \right\} \tag{5}$$

where $\zeta \equiv 1 - z/f$. The above was calculated using the ABCD matrix description of a thin lens of focal distance *f*. The intensity distribution after the lens can then be estimated by:

$$I'(\mathbf{r},z) \propto \frac{1}{\zeta^2} \left\{ \left|\Psi_o\left[\frac{\mathbf{r}}{\zeta}, \frac{z}{\zeta}\right]\right|^2 + \left|\Psi_o\left[\frac{\mathbf{r}}{\zeta}, \frac{-z}{\zeta}\right]\right|^2 + 2\mathrm{Re}\left[\Psi_o\left[\frac{\mathbf{r}}{\zeta}, \frac{z}{\zeta}\right] \Psi_o\left[\frac{\mathbf{r}}{\zeta}, \frac{-z}{\zeta}\right]\right] \right\} \tag{6}$$

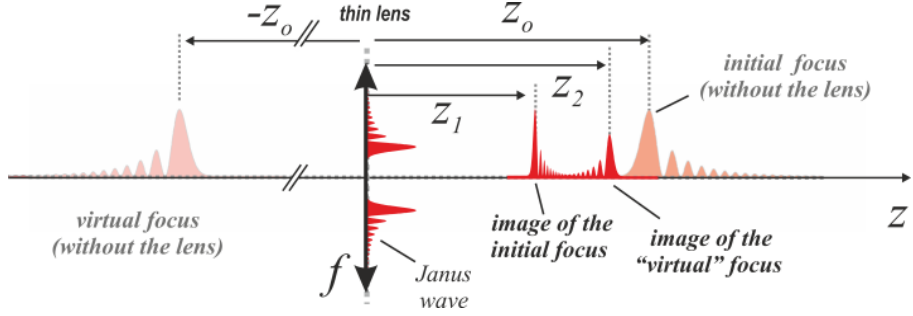

**Fig. 2** Focusing a Janus wave with a thin lens. Without the presence of the lens each intensity peak (focus) is accompanied by a symmetrically positioned "virtual" intensity peak. The focusing action of the lens creates two opposite facing focal distributions.

As expected the resulting intensity distribution is the result of the interference of the two waves. Interestingly, a local maximum of the wave distribution $\Psi_o(\mathbf{r}, z)$ at position $(\mathbf{r}_o, z_o)$ will be imaged after the lens at two positions $(\mathbf{r}_1, z_1)$, $(\mathbf{r}_2, z_2)$:

$$\left. \begin{array}{l} z_o \equiv \dfrac{z_1}{\zeta_1}, \mathbf{r}_o \equiv \dfrac{\mathbf{r}_1}{\zeta_1} \\[6pt] z_o \equiv \dfrac{-z_2}{\zeta_2}, \mathbf{r}_o \equiv \dfrac{\mathbf{r}_2}{\zeta_2} \end{array} \right\} \Rightarrow \begin{array}{l} -\dfrac{1}{z_o} + \dfrac{1}{z_1} = \dfrac{1}{f}, \mathbf{r}_1 = M_T(-z_o) \cdot \mathbf{r}_o \\[6pt] \dfrac{1}{z_o} + \dfrac{1}{z_2} = \dfrac{1}{f}, \mathbf{r}_2 = M_T(z_o) \cdot \mathbf{r}_o \end{array}$$

(7)

where $\zeta_i = 1 - z_i/f$, and $M_T(z) = (1 - z/f)^{-1}$ is the transverse magnification of the lens. Although the accurate location of the intensity maxima from eq. (6) involves a complex analysis, the positions of the twin images, as given by eq. (7), are correct as long as $z_o \neq 0$, $|1 - z_i/f| > 0.1$. Counterintuitively, classical imaging formulas can be used to predict the maxima positions if one takes into account that the peak of the $\Psi_o(\mathbf{r}, z)$ wave can be treated as an imaginary object to the lens ($-z_o$) while the peak of the "conjugate" $\Psi_o^*(\mathbf{r}, -z)$ wave can be treated as a real object ($z_o$).

Furthermore, the transverse position of the peaks is predicted by taking into account the transverse magnification $M_T$ for each position of the object. It is not difficult to see that the two foci will exhibit opposite signs of $M_T$ for $z_o > f$. As shown schematically in Fig. 2 without the presence of the lens each initial intensity peak (focus) of the wave is accompanied by a symmetrically positioned virtual focus. The focusing action of the lens leads to the creation of two opposite facing focal distributions located at positions $z_1, z_2$ that can be estimated using eq. (7). The two foci are distinct as long as the twin waves, $\Psi_o(\mathbf{r}, z)$ and $\Psi_o^*(\mathbf{r}, -z)$ do not have overlapping maxima along z. It is easy to show that overlapping maxima can occur only on the symmetry plane $z_o = 0$, a case already excluded from the definition of JWs. Thus a JW, in contrast to the broader category of CWs, will always exhibit two foci under the action of a lens.

The validity of our analysis can be confirmed by applying it in the case of colliding 1D Airy beams[21]. These beams are the 1D equivalent of 2D rotationally symmetric ring-Airy beams [7, 8, 22] and are consisting of two 1D Airy beams that propagate along symmetric parabolic trajectories.

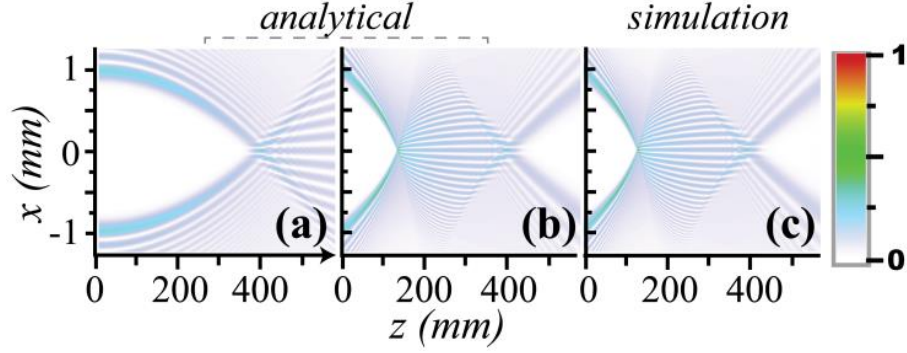

**Fig. 3** Intensity distribution of 1D colliding Airy beams. Comparison of analytical prediction with numerical simulation results (a) Free propagation, (b), (c) propagation after being focused by a thin lens. (parameters $x_o$ =1 mm, $w = 84.25$ μm, $\alpha = 0.05$, $\lambda = 800$ nm, $f = 200$ mm)

The two beams coherently interfere generating a focal point on the propagation axis. In this case the beam propagation can be analytically described [4] by:

$$U_{1D}(x,z) = \Phi(x,z) + \Phi(-x,z),$$

$$\Phi(x,z) = Ai\left(\frac{x_o+x}{w} - \frac{z^2}{4k^2w^4} + i\frac{az}{kw^4}\right) e^{a\frac{x_o+x}{w} - \frac{az^2}{4k^2w^4} + i\left(\frac{a^2z}{2kw^2} + \frac{x_o+x}{2kw^3}z - \frac{z^3}{12k^3w^6}\right)}$$

(8)

where $x$ is the spatial coordinate, $x_o$ and $w$ are respectively the primary lobe position and width parameters and $k$ is the wavenumber. It is clear that at $z = 0$ the wave is real valued. In this case, following eq. (1), the wave can be decomposed to the interference of two waves, $\Psi_o(x,z)$ and a "conjugate" $\Psi_o^*(x,-z)$ by simply setting: $\Psi_o(x,z) = U_{1D}(x,z)/2$.

As shown in Fig. 3(a) the symmetric Airy beams propagate following a parabolic trajectory generating a focus at $z_o \cong 400$ mm. When this beam is focused by a lens, the resulting intensity distribution is a double foci image together with an interference pattern between the $\Psi_o, \Psi_o^*$ conjugate waves. Combining now Eqs. (8) and (5), adapted for 1D propagation, we can analytically formulate the wave distribution after the insertion of a thin lens at $z = 0$. The analytically estimated intensity distribution (from eqs. (8), (5)) is shown together with numerical simulation results in Figs. 3(b),(c), and as one can clearly see the agreement is excellent. Furthermore, the two foci are located at $z_1 = 133$ *mm* and $z_2 = 400$ *mm* as predicted by eq. (7).

**Experimental results**

For the experimental demonstration of JWs we have chosen to work with ring- Airy beams [7]. The experimental setup used to study this effect is shown in Fig. 4. A cw laser diode emitting at 800 *nm* was used as a source. In order to generate the ring-Airy beam we followed a Fourier Transform approach described in detail in [8]. In brief, we first used a phase only reflecting spatial light modulator (SLM) in order to modulate the phase of a Gaussian beam. The modulated beam was then Fourier Transformed by a lens. The ring-Airy distribution is generated in the Fourier transform plane of the lens after blocking the zero order diffraction. We have tuned the parameters of the ring-Airy beam so that the abrupt autofocus position was at $z_o = 400$ *mm*.

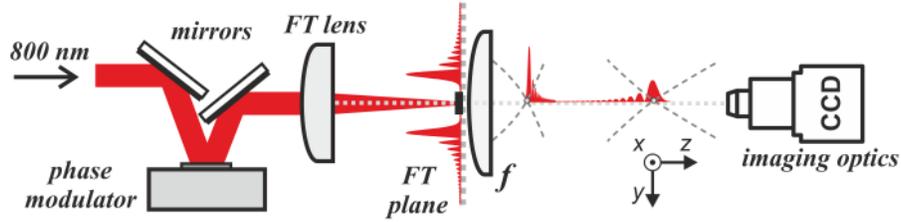

**Fig. 4** Experimental setup. FT: Fourier Transform. FT lens (500 *mm*). focusing lens *f*: 200 *mm*, (the beam propagates from left to right)

The transverse intensity distribution as the beam propagated was imaged using a compact microscope system (0.2 NA) and a linear CCD camera (14 bit).

We firstly studied the ring-Airy beam propagation without the presence of a focusing lens and then we focused this beam by placing a plano-convex lens of focal distance $f = 200$ *mm* close to the ring –Airy generation plane. The $I(x,y,z)$ intensity profile of the beam was retrieved by merging 2D transverse (x,y) images captured at various positions along the propagation axis (*z*).

An *x-z* cross section of the intensity profile is shown in Fig. 5 together with numerical simulations of the paraxial wave equation. As expected from our former analysis for JW waves the focused ring-Airy beams form two focal regions. The "real" and "virtual" waves produce two foci at different positions and the positions of the two foci agree well, within 2%, to the theoretical predictions of eq. (7). Finally, as can be seen in Fig. 5 the agreement between the experimental measurements and the simulations is remarkable.

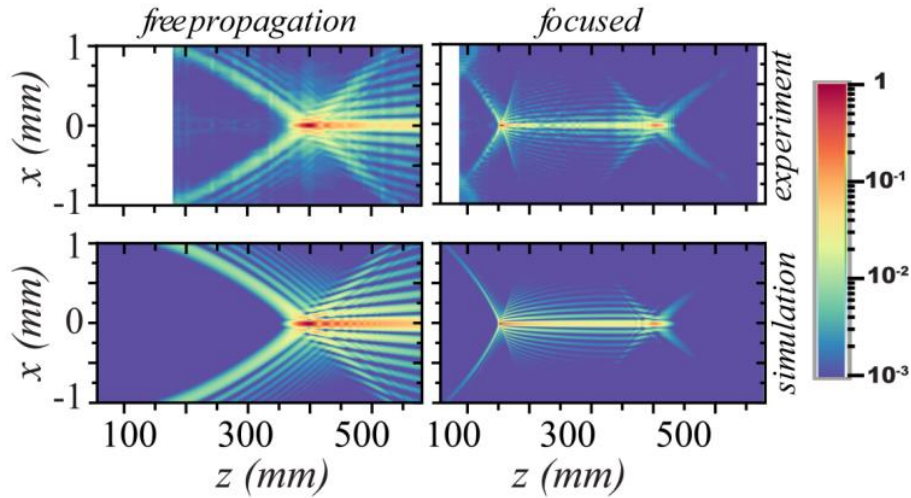

**Fig. 5** Cross sectional intensity profiles $I(x,0,z)$ of ring-Airy beams. Experimental measurements (top row), simulation (bottom row). The simulations are performed assuming a thin lens located at position z=0. Free propagation (left column), focusing by a 200 mm lens (right column)

**Conclusions**

We have introduced a family of waves, called "Janus waves", whose propagation can be decomposed to "real" and "virtual" components, conjugate to each other under inversion of the propagation direction. From the theoretical point of view, we have established simple criteria for a wave to belong to the Janus wave family. On the other hand, from the experimental point of view Janus waves can be easily identified by the appearance of two focal regions after the action of a focusing lens.

We have derived analytic formulas for the intensity distribution after the action of a focusing lens, while interestingly, simple imaging formulas can be used to predict the intensity peak positions. Finally, this exciting behavior is demonstrated both experimentally and by simulations for the case accelerating ring-Airy beams. Due to the equivalence to holography, our approach can be generalized to all waves that rely on diffraction to generate an intense focal point. We expect that Janus waves will find exciting applications both in linear and nonlinear optics. The possibility to engineer symmetric foci distributions can for instance impact optical trapping applications, or the controlled deposition of high laser powers at remote locations.

**Funding.** QNRF project No. NPRP9-383-1-083, Laserlab-Europe (EU-FP7 284464)